# CONSTRUCTION OF SCHRÖDINGER, PAULI AND DIRAC EQUATIONS FROM VLASOV EQUATION IN CASE OF LORENTZ GAUGE

**E.E. Perepelkin[a,c,d], B.I. Sadovnikov[a], N.G. Inozemtseva[b,c], M.V. Klimenko[a]**

[a] *Faculty of Physics, Lomonosov Moscow State University, Moscow, 119991 Russia*
[b] *Moscow Technical University of Communications and Informatics, Moscow, 123423 Russia*
[c] *Dubna State University, Moscow region, Dubna, 141980 Russia*
[d] *Joint Institute for Nuclear Research, Moscow region, Dubna, 141980 Russia*

**Abstract**

On the basis of the first principle – the law of probability conservation and the Helmholtz decomposition theorem the authors have succeeded to construct the Schrödinger, Pauli, Dirac equation, the Hamilton–Jacobi equation and the Maxwell equations. The approach described in this paper makes it possible to naturally connect the classical and quantum systems.

**Introduction**

In the middle of the 20th century Vlasov invented the first principle – the probability conservation law in the generalized phase space (GPS) $\Omega_\infty$ [1]:

$$\frac{\partial}{\partial t}f_\infty + \text{div}_\xi\left(f_\infty \vec{u}_\xi\right) = 0, \tag{i.1}$$

where $\text{div}_\xi \overset{\text{def}}{=} \text{div}_r + \text{div}_v + \text{div}_{\dot v} + \text{div}_{\ddot v} + ...$; $f_\infty(\vec\xi, t)$ is a distribution function or probability density function; $\vec\xi = \{\vec r, \vec v, \dot{\vec v}, \ddot{\vec v}, \dddot{\vec v}, ...\}^T \in \Omega_\infty$; $\vec{u}_\xi \overset{\text{def}}{=} \{\vec v, \dot{\vec v}, \ddot{\vec v}, ...\}^T$ is a generalized velocity. For each point $\vec\xi_0 = \{\vec r_0, \vec v_0, \dot{\vec v}_0, ...\}^T \in \Omega_\infty$ there is a corresponding generalized phase trajectory [2]:

$$\vec\xi(t) = e^{t\hat D}\vec\xi_0 = M_\infty(t)\vec\xi_0, \tag{i.2}$$

$$\vec{u}_\xi = e^{t\hat D}\vec{u}_{\xi_0}, \quad \dot{\vec{u}}_\xi = e^{t\hat D}\dot{\vec{u}}_{\xi_0}, \quad \ddot{\vec{u}}_\xi = e^{t\hat D}\ddot{\vec{u}}_{\xi_0}, ... \tag{i.3}$$

$$M_{N_a}(t) \overset{\text{def}}{=} \begin{pmatrix} 1 & t & t^2/2 & t^3/3!.. \\ 0 & 1 & t & t^2/2... \\ 0 & 0 & 1 & t... \\ ... & ... & ... & .... \\ 0 & ... & ... & 0 \quad 1 \end{pmatrix}, \quad M_\infty = \lim_{N_a \to +\infty} M_{N_a}, \quad \det M_\infty = 1, \tag{1.4}$$

where $M_\infty$ is the Taylor evolutional matrix in GPS, and $M_{N_a}$ is an evolutional matrix $N_a \times N_a$ with order of approximation or averaging $N_a$; $\hat D$ is the differentiation operator along the generalized phase trajectory $\vec{u}_\xi \overset{\text{def}}{=} \hat D \vec\xi \overset{\text{def}}{=} \{\vec v, \dot{\vec v}, \ddot{\vec v}, ...\}^T$. Note that point $\vec\xi_0 \in \Omega_\infty$ defines the one-parameter group of the Lie transformations.

Continual integration of equation (i.1) over phase subspaces $\Omega^s : \Omega_\infty = \Omega^1 \times \Omega^2 \times ...$ leads to the infinite self-linking chain of the Vlasov equations. The chain of equations can be written in a compact and physically clear form [3]:

$$\hat{\pi}_{1...n} S^{1...n} = -Q_{1...n}, \quad n \in \mathbb{N}, \tag{i.5}$$

where

$$\hat{\pi}_{1...n} \stackrel{def}{=} \frac{\partial}{\partial t} + \vec{v} \cdot \nabla_r + \dot{\vec{v}} \cdot \nabla_v + ... + \left\langle \vec{\xi}^{n+1} \right\rangle_{1...n} \cdot \nabla_{\xi^n}, \tag{i.6}$$

$$S^{1,..,n} \stackrel{def}{=} \operatorname{Ln} f^{1,..,n}, \quad Q_{1,..,n} \stackrel{def}{=} \operatorname{div}_{\xi^n} \left\langle \vec{\xi}^{n+1} \right\rangle_{1,..,n}, \tag{i.7}$$

$$f^{1...n} \stackrel{def}{=} \prod_{s=n+1}^{+\infty} \int_{\Omega^s} f_\infty(\vec{\xi}, t) d^3 \xi^s, \quad \left\langle \vec{\xi}^{n+1} \right\rangle_{1...n} \stackrel{def}{=} \frac{1}{f^{1...n}} \int_{\Omega^{n+1}} f_\infty(\vec{\xi}, t) \vec{\xi}^{n+1} d^3 \xi^{n+1} \prod_{s=n+2}^{+\infty} \int_{\Omega^s} d^3 \xi^s. \tag{i.8}$$

Operator (i.6) corresponds to the total derivative with respect to time in phase subspace of $n$ kinematical values. Values $Q_{1,..,n}$ (i.7) determine the density of dissipation sources. As a result, from equations (i.5) it follows that the change in the probability density along the phase trajectory (i.2) is equal to the sources of dissipation. The first Vlasov equation ($n=1$) is known as the continuity equation in continuum mechanics and the field theory. If there are no sources of dissipation $Q_1 = \operatorname{div}_r \left\langle \vec{v} \right\rangle_1 = 0$, then the probability density $f^1 = const$ along the trajectory of motion (i.2) in the coordinate space. The second equation ($n=2$) is used in plasma physics, astrophysics, solid state physics, statistical physics [4-10] and is known as the Vlasov equation. If there are no sources of dissipation $Q_{1,2} = \operatorname{div}_v \left\langle \dot{\vec{v}} \right\rangle_{1,2}, Q_{1,2} = 0$, then the second Vlasov equation turns into the Liouville equation [11]. Therefore, density function $f^{1,2} = const$ along the phase trajectory (i.2).

To solve the Vlasov chain (i.5) it is necessary to cut this chain off at a certain equation which determines the amount of kinematic information $n$ about the system. To do this we need an approximation of the field $\left\langle \vec{\xi}^{n+1} \right\rangle_{1...n}$ [18]. In this paper, we consider the chain cut-off (i.5) at the first equation ($n=1$) by expressing the field $\left\langle \vec{v} \right\rangle_1$ via the Helmholtz decomposition into a potential and vortex component. This approach was implemented for the Coulomb gauge earlier in [13]. Now we consider an extended case, namely the Lorentz gauge. The representation of the Helmholtz decomposition in terms of hypercomplex functions made it possible to obtain analogues of the Schrödinger, Pauli and Dirac equations for electromagnetic systems from the first Vlasov equation.

The work has the following structure. In section 1 from the first equation of the chain the Schrödinger equation taking into account the Lorentz gauge, the equation of motion of a charged particle in an electromagnetic field as well as the Hamilton-Jacobi equation are obtained. The concept of the self-consistent system is introduced, which makes it possible to construct the Maxwell equations both for external fields and for probabilistic (quantum) fields of a particle. Paragraph 2 is devoted to obtaining the Pauli equation and the Dirac equation from the first Vlasov equation by increasing the amount of kinematic information about the quantum system. The basis is the Helmholtz decomposition (§1) for the vector field of probability flow, in which the transition is made from a complex function to a spinor and a bispinor. The Conclusions paragraph contains a description of the main work results. The proofs of the theorems are given in Appendix A and B.

**§1 Schrödinger equation**

The Schrödinger equation was known to be originally obtained using a phenomenological method [12]. In [13] the Schrödinger equation, the Hamilton-Jacobi equation, the equation of motion of a charged particle in an electromagnetic field and the Maxwell system of equations are obtained from the first Vlasov equation (i.5), ($n=1$), which is based on the first principle of the probability conservation law (i.1).

When obtaining the Schrödinger equation for a scalar particle in an electromagnetic field [13], the Coulomb gauge was used $\text{div}_r \vec{A} = 0$. Note that an analogue result may be obtained for the Lorentz gauge $\text{div}_r \vec{A} = \chi$, where $qc^2 \chi = -\partial U/\partial t$. For clarity of notation, we omit the indices of functions and fields indicating the type of phase subspace.

The first Vlasov equation (i.5), ($n=1$) contains two unknown functions $f^1 = f(\vec{r},t)$ and $\langle \vec{v} \rangle_1 = \langle \vec{v} \rangle(r,t)$. Such an equation is possible to be solved in two cases: either function $\langle \vec{v} \rangle$ or $f$ is known, or there is a mathematical relationship between functions $\langle \vec{v} \rangle$ and $f$. From a physical point of view, the Vlasov equation describes a certain physical object, therefore functions $\langle \vec{v} \rangle$ and $f$ must correspond to the same object. Therefore, it is logical to consider the second case, when there is a mathematical relationship between $\langle \vec{v} \rangle$ and $f$ in addition to the Vlasov equation. As the simplest mathematical object that combines two quantities, one can take a complex function. A complex function is known to have two components: a real and an imaginary part. On the complex plane it is convenient to work with the Euler representation of a complex number in the form of its modulus and phase.

Based on the above, we will use complex function $\Psi \in \mathbb{C}$ as a representation that combines two real functions $f$ and $\langle \vec{v} \rangle$ included in the first Vlasov equation. Let us set the modulus of complex number $\Psi$ equal to $|\Psi|^2 = f \geq 0$. The positivity property of probability density $f$ will be satisfied automatically. We relate scalar phase $\varphi = \arg \Psi$ with vector field $\langle \vec{v} \rangle$ via the Helmholtz theorem [14] on the decomposition of the vector field (i.8) ($n=1$) into vortex $\vec{A}_\Psi$ and potential $\nabla_r \Phi$ fields:

$$\langle \vec{v} \rangle(\vec{r},t) = -\alpha \nabla_r \Phi(\vec{r},t) + \gamma \vec{A}_\Psi(\vec{r},t), \ \text{div}_r \vec{A}_\Psi = \chi, \qquad (1.1)$$

where $\alpha, \gamma$ are some constant values, and $\chi$ is some free function. Let us transform expansion (1.1)

$$\langle \vec{v} \rangle = i^2 \alpha \nabla_r \Phi + \gamma \vec{A}_\Psi = i\alpha \nabla_r (0 + i\Phi) + \gamma \vec{A}_\Psi = i\alpha \nabla_r \left( \ln \left| \frac{\Psi}{\Psi^*} \right| + i\Phi \right) + \gamma \vec{A}_\Psi, \qquad (1.2)$$

$$\frac{\Psi}{\Psi^*} = e^{i2\varphi} \Rightarrow \text{Arg}\left[ \frac{\Psi}{\Psi^*} \right] = 2\varphi(\vec{r},t) + 2\pi k \stackrel{\text{def}}{=} \Phi(\vec{r},t), \ k \in \mathbb{Z}. \qquad (1.3)$$

Taking into account definition (1.3), the representation (1.2) of average velocity takes the form:

$$\langle \vec{v} \rangle(\vec{r},t) = i\alpha \nabla_r \left( \ln \left| \frac{\Psi}{\Psi^*} \right| + i \text{Arg}\left[ \frac{\Psi}{\Psi^*} \right] \right) + \gamma \vec{A}_\Psi = i\alpha \nabla_r \text{Ln}\left[ \frac{\Psi}{\Psi^*} \right] + \gamma \vec{A}_\Psi,$$

or

$$\vec{J}(\vec{r},t) \stackrel{\text{def}}{=} f(\vec{r},t)\langle\vec{v}\rangle(\vec{r},t) = i\alpha\left[\Psi^*\nabla_r\Psi - \Psi\nabla_r\Psi^*\right] + \gamma\Psi^*\vec{A}_\Psi\Psi,$$

$$J^k = \Psi^*\left(-\alpha\partial^k\Phi + \gamma A_\Psi^k\right)\Psi, \quad k=1...3, \qquad (1.4)$$

where $\vec{J}(\vec{r},t)$ is the current density and it is taking into account that $|\Psi|^2 = f$.

**Definition 1** *Representation (1.4) of vector field $\langle\vec{v}\rangle(\vec{r},t)$ or field $\vec{J}(\vec{r},t)$ for $\Psi \in \mathbb{C}$ will be called the Helmholtz $\Psi\mathbb{C}$– decomposition with gauge $\text{div}_r\,\vec{A}_\Psi = \chi$.*

**Theorem 1** *If the vector field $\langle\vec{v}\rangle(\vec{r},t)$ of the probability flow admits a Helmholtz $\Psi\mathbb{C}$– decomposition with gauge $\text{div}_r\,\vec{A}_\Psi = \chi$, then the first Vlasov equation for function $f = \Psi\Psi^*$ becomes the $\Psi$-equation:*

$$\frac{i}{\beta}\frac{\partial\Psi}{\partial t} = -\alpha\beta\left(\hat{p} - \frac{\gamma}{2\alpha\beta}\vec{A}_\Psi\right)^2\Psi + U\Psi, \qquad (1.5)$$

*where $\hat{p} \stackrel{\text{def}}{=} -\frac{i}{\beta}\nabla_r$ and $\beta \neq 0$, $\beta \in \mathbb{R}$, is a constant value; $U(\vec{r},t) \in \mathbb{R}$ is some function.*

The proof of Theorem 1 is given in Appendix A.

If constant values $\alpha,\beta,\gamma$ are chosen as:

$$\alpha = -\frac{\hbar}{2m}, \quad \beta = \frac{1}{\hbar}, \quad \gamma = -\frac{q}{m}, \qquad (1.6)$$

then equation (1.5) will take the form of the Schrödinger equation for a scalar particle in an electromagnetic field

$$i\hbar\frac{\partial\Psi}{\partial t} = \frac{1}{2m}\left(\hat{p} - q\vec{A}_\Psi\right)^2\Psi + U\Psi, \qquad (1.7)$$

where $\hat{p} = -i\hbar\nabla_r$ corresponds to the momentum operator. Note that, despite the presence of free function $\chi$, the equation (1.5)/(1.7) does not depend on it explicitly. Let us show that function $U$ corresponds to the potential energy.

**Theorem 2** *Function $U(\vec{r},t) \in \mathbb{R}$ from equation (1.5) corresponds the equation:*

$$-\frac{1}{\beta}\frac{\partial\varphi}{\partial t} = -\frac{1}{4\alpha\beta}|\langle\vec{v}\rangle|^2 + V \stackrel{\text{det}}{=} H, \qquad (1.8)$$

*where*

$$V = U + Q, \quad Q = \frac{\alpha}{\beta}\frac{\Delta_r|\Psi|}{|\Psi|}, \qquad (1.9)$$

and vector field $\langle\vec{v}\rangle(\vec{r},t)$ of the probability flow satisfies the equation of motion:

$$\hat{\pi}_1\langle\vec{v}\rangle = \frac{d}{dt}\langle\vec{v}\rangle = -\gamma\left(\vec{E}_\Psi + \langle\vec{v}\rangle\times\vec{B}_\Psi\right), \qquad (1.10)$$

where

$$\vec{E}_\Psi \stackrel{def}{=} -\frac{\partial \vec{A}_\Psi}{\partial t} - \frac{2\alpha\beta}{\gamma}\nabla_r V, \quad \vec{B}_\Psi \stackrel{def}{=} \mathrm{curl}_r \vec{A}_\Psi. \qquad (1.11)$$

The proof of Theorem 2 is given in Appendix A.

When replacing (1.6), equation (1.8) becomes the Hamilton-Jacobi equation $-\hbar\frac{\partial\varphi}{\partial t} = \frac{m}{2}|\langle\vec{v}\rangle|^2 + V = H$, and expression (1.9) determines the quantum potential $Q = -\frac{\hbar^2}{2m}\frac{\Delta_r|\Psi|}{|\Psi|}$ from the de Broglie-Bohm «pilot-wave» theory [15-17]. Note that potential V in the Hamilton function H differs from potential $U$ in the Schrödinger equation by the value of the quantum potential Q. In classical limit $\hbar \ll 1$ we can assume that $V \approx U$. Value $\hbar\varphi = S$ determines the action.

**Definition 2** *We will call the relationship between vortex field $\vec{A}_\Psi$ component (1.4) and scalar potential* V *(1.9) using the equation*

$$\mathrm{div}_r \vec{A}_\Psi + \frac{2\alpha\beta}{\gamma}\frac{1}{c^2}\frac{\partial V}{\partial t} = 0, \qquad (1.12)$$

*Lorentz* $\Psi$ *-gauge.*

Without loss of generality, vortex field $\vec{A}_\Psi$ can be represented as the sum of classical vector potential $\vec{A}$ and some «quantum» vector potential $\vec{A}_Q$, that is, (1.11)

$$\vec{A}_\Psi \stackrel{det}{=} \vec{A} + \vec{A}_Q, \quad \vec{B}_\Psi = \mathrm{curl}_r \vec{A} + \mathrm{curl}_r \vec{A}_Q = \vec{B} + \vec{B}_Q. \qquad (1.13)$$

In the classical approximation ($\hbar \ll 1$) for $Q = 0$ and $\vec{A}_Q = \vec{\theta}$ according to (1.11), (1.12) and (1.6), the Lorentz $\Psi$-gauge (1.12) turns into the usual Lorentz gauge

$$\mathrm{div}_r \vec{A} + \frac{1}{qc^2}\frac{\partial U}{\partial t} = 0. \qquad (1.14)$$

In the general case, when taking into account quantum potential Q and the vector potential $\vec{A}_Q$, $\Psi$−gauge (1.12) splits into Lorentz gauge (1.14) and quantum gauge

$$\mathrm{div}_r \vec{A}_Q + \frac{1}{qc^2}\frac{\partial Q}{\partial t} = 0. \qquad (1.15)$$

Electric field $\vec{E}_\Psi$ (1.11) takes the form:

$$\vec{E}_\Psi = -\frac{\partial \vec{A}}{\partial t} - \frac{1}{q}\nabla_r U - \frac{\partial \vec{A}_Q}{\partial t} - \frac{1}{q}\nabla_r Q = \vec{E} + \vec{E}_Q, \qquad (1.16)$$

where $\vec{E}, \vec{B}$ are classical electromagnetic fields; and $\vec{E}_Q, \vec{B}_Q$ are electromagnetic fields caused by the probability distributions of the quantum system. The charge $\rho$ and current $\vec{J}$ densities are determined for classical fields $\vec{E}, \vec{B}$. For a quantum system located in external classical fields, according to the uncertainty principle, charge density $qf$ (related to the coordinate) and current density $qf\vec{v}$ (related to velocity $\vec{v}$) are not determined. The indicated densities are not classical electrodynamics characteristics, they are probabilistic by nature. For example, field $\vec{E}_Q$ (1.16) includes the quantum potential Q, expressed in terms of density $\sqrt{f} = |\Psi|$ as (1.9). Dependence (1.9), in the general case, does not agree with the Coulomb representation $\Delta_r Q = -qf/\varepsilon_0$.

**Definition 3** *We will say that the system is self-consistent if the condition (1.17) is satisfied for* $\rho_\Psi \stackrel{def}{=} \varepsilon_0 \operatorname{div}_r \vec{E}_\Psi \stackrel{def}{=} \operatorname{div}_r \vec{D}_\Psi$

$$\rho_\Psi = qf. \qquad (1.17)$$

**Lemma 1** *For a self-consistent system (1.17), the following field equations are valid:*

$$\operatorname{div}_r \vec{D}_\Psi = \rho_\Psi, \quad \operatorname{div}_r \vec{B}_\Psi = 0, \qquad (1.18)$$

$$\operatorname{curl}_r \vec{E}_\Psi = -\frac{\partial \vec{B}_\Psi}{\partial t}, \quad \frac{\partial}{\partial t}\vec{D}_\Psi + \vec{J}_\Psi = \operatorname{curl}_r \vec{H}_\Psi, \qquad (1.19)$$

*where* $\vec{J}_\Psi \stackrel{def}{=} \rho_\Psi \langle\vec{v}\rangle$, $\vec{H}_\Psi$ *are some field, for example,* $\mu_0 \vec{H}_\Psi \stackrel{def}{=} \vec{B}_\Psi$. *In this case, the Lorentz* $\Psi-$ *gauge condition (1.12) brings the system of equations (1.18)-(1.19) to the form:*

$$\Box\varphi_\Psi = \frac{\rho_\Psi}{\varepsilon_0}, \quad \Box\vec{A}_\Psi = \mu_0 \vec{J}_\Psi, \qquad (1.20)$$

$$\varphi_\Psi = \varphi + \varphi_Q, \quad \vec{J}_\Psi = \vec{J} + \vec{J}_Q, \qquad (1.21)$$

$$V \stackrel{def}{=} q\varphi_\Psi, \quad Q \stackrel{def}{=} q\varphi_Q, \quad U = q\varphi, \qquad \rho_\Psi = \rho + \rho_Q, \qquad (1.22)$$

*where* $\rho_Q \stackrel{def}{=} \varepsilon_0 \operatorname{div}_r \vec{E}_Q \stackrel{def}{=} \operatorname{div}_r \vec{D}_Q$.

The proof of Lemma 1 is given in Appendix A.

**Remark 1** In the classical limit ($\hbar \ll 1$) at $Q = 0$ and $\vec{A}_Q = \vec{0}$ equations (1.18)-(1.19) for a self-consistent system transform into the well-known Maxwell equations for classical electromagnetic fields $\vec{E}, \vec{B}$:

$$\text{div}_r \, \vec{D} = \rho, \quad \text{div}_r \, \vec{B} = 0, \tag{1.23}$$

$$\text{curl}_r \, \vec{E} = -\frac{\partial \vec{B}}{\partial t}, \qquad \frac{\partial}{\partial t}\vec{D} + \vec{J} = \text{curl}_r \, \vec{H}, \tag{1.24}$$

where $\vec{J} = \rho \langle \vec{v} \rangle$ and $\mu_0 \vec{H} = \vec{B}$. In this case, the Lorentz gauge (1.14) can be performed. In the general case, when taking into account «quantum» fields $\vec{E}_Q, \vec{B}_Q$, and having classical Maxwell equations (1.23)-(1.24) being true for fields $\vec{E}, \vec{B}$, we obtain an analogue of the Maxwell equations for fields $\vec{E}_Q, \vec{B}_Q$:

$$\text{div}_r \, \vec{D}_Q = \rho_Q, \quad \text{div}_r \, \vec{B}_Q = 0, \tag{1.25}$$

$$\text{curl}_r \, \vec{E}_Q = -\frac{\partial \vec{B}_Q}{\partial t}, \qquad \frac{\partial}{\partial t}\vec{D}_Q + \vec{J}_Q = \text{curl}_r \, \vec{H}_Q, \tag{1.26}$$

where $\vec{J}_Q = \rho_Q \langle \vec{v} \rangle$ and $\mu_0 \vec{H}_Q = \vec{B}_Q$. In this case, for equations (1.25)-(1.26) quantum gauge (1.15) can be performed.

**Remark 2** Note that field equations (1.18)-(1.19) for self-consistent systems are obtained only on the basis of the first Vlasov equation, which is a consequence of the first principle – the probability conservation law. Equations (1.18)-(1.19) in the form (1.20) are representable through 4-potential $A^\mu$, $\mu = 0...3$

$$\Box A^\mu = \mu_0 J^\mu, \tag{1.27}$$

$$A^\mu = \left( \frac{\varphi_\Psi}{c}, \vec{A}_\Psi \right), \qquad J^\mu = \left( c\rho_\Psi, \vec{J}_\Psi \right),$$

where $\Box = \partial^\mu \partial_\mu$, $\partial_\mu = \dfrac{\partial}{\partial x^\mu}$, $\partial^\mu = g^{\mu\nu} \partial_\nu$. Metric tensor $g^{\mu\nu} = (+,-,-,-)$ corresponds to the d'Alembert operator (1.27). Equation (1.27) is invariant under the Lorentz transformation, thus it is a relativistic equation. The problem of integration (averaging) over the velocity space (i.8) is solved rather by momentum than velocity. Indeed, with $v \to c$ momentum $p \to \infty$. In this case, equation (1.27) is uniquely related to the «Schrödinger equation» (1.7). Note that equation (1.7) is not invariant under Lorentz transformations, since it contains derivatives of different orders.

Equation (1.7) is not exactly the Schrödinger equation. In the general case, there is a vector potential $\vec{A}_\Psi$ that differs, according to (1.13), from the classical vector potential $\vec{A}$. A similar remark concerns the field equation itself (1.27), in which, in addition to vector potential $\vec{A}_\Psi$, there is a scalar potential $\varphi_\Psi$ (1.21) associated with the quantum potential Q (1.9). In the nonrelativistic approximation, only external fields $\vec{E}, \vec{B}$ that satisfy the Maxwell equations remain, and the relation with particle fields (quantum fields) $\vec{E}_Q, \vec{B}_Q$ disappears. The classical Schrödinger equation does not contain the particle's own fields. There are only external potentials $\vec{A}$ and $U = q\varphi$. The influence of the particle's own «field», for example, in the form of spin, is manifested in the equations of Pauli and Dirac (see §2).

Recall that according to the Helmholtz $\Psi\mathbb{C}$ – decomposition (1.4), the velocity of probability flow $\langle \vec{v} \rangle$ is related to phase $\varphi$ of the wave function $\Psi$. Probabilistic fields $\vec{E}_Q, \vec{B}_Q$ provide probabilistic solutions to equation of motion (1.10) for flow $\langle \vec{v} \rangle$, and therefore also for phase $\varphi$, the incursion of which is present in Feynman's path integrals.

**§2 Pauli and Dirac equations**

Let us consider a possible option for taking into account additional kinematic information (vortex field $\vec{A}_Q$) of the particle itself located in external fields. When obtaining an analogue of the Schrödinger equation (1.7), the system was described by a complex function $\Psi \in \mathbb{C}$. Complex function $\Psi = |\Psi| \exp(i\varphi)$ contains two free parameters: modulus $|\Psi|$ and phase $\varphi$. These two parameters were used to describe two unknown functions in the Vlasov equation (i.5), ($n=1$): probability density $f = |\Psi|^2$ and probability flow velocity $\langle \vec{v} \rangle = -2\alpha \nabla_r \varphi + \gamma \vec{A}_\Psi$. At the same time, in order to describe the vortex field, it was necessary to introduce another free parameter $\vec{A}_\Psi$, which, according to (1.13), in addition to the vector-potential $\vec{A}$ (external magnetic field), contains the probabilistic vortex field of the particle $\vec{A}_Q$. From equation (1.7) it is impossible to explicitly obtain the form of field $\vec{A}_Q$, which is included there as a parameter. Of course, field $\vec{A}_Q$ satisfies the gauge expression (1.15), but according to the Helmholtz theorem, the field cannot be reconstructed from divergence ($\text{div}_r$) only [18]. Field $\vec{A}_Q$ can be found from the system of equations (1.25)-(1.26), but for this it is necessary to know probability density $\rho_Q$ and current $\vec{J}_Q$, which is expressed through field $\vec{A}_Q$ itself.

Thus, usual complex function $\Psi \in \mathbb{C}$ does not have a sufficient number of free parameters to describe the particle's own kinematic characteristics. To solve this problem, we increase the number of free parameters by considering a 2D complex space $\mathbb{C}^2$ with elements $\psi$ (spinors):

$$\psi = \begin{pmatrix} \psi_1 \\ \psi_2 \end{pmatrix} \in \mathbb{C}^2, \quad \psi^\dagger = \begin{pmatrix} \psi_1^* & \psi_2^* \end{pmatrix} = \left(\psi^*\right)^T, \quad |\psi|^2 = \psi^\dagger \psi, \quad \psi_1, \psi_2 \in \mathbb{C}. \qquad (2.1)$$

Spinor $\psi$, unlike complex function $\Psi$, has four instead of two free parameters. By analogy with the Helmholtz $\Psi\mathbb{C}$–decomposition (1.4) for the function $\Psi \in \mathbb{C}$, we define the decomposition for spinor $\psi \in \mathbb{C}^2$.

**Definition 4** *Let us denote the representation of current density field $\vec{J}(\vec{r},t)$ for $\psi \in \mathbb{C}^2$*

$$\vec{J} = i\alpha \left( \psi^\dagger \nabla_r \psi - \psi^T \nabla_r \psi^* \right) + \gamma \psi^\dagger \vec{A} \psi, \qquad (2.2)$$

*as the Helmholtz $\psi\mathbb{C}^2$ – decomposition with gauge $\text{div}_r \vec{A} = \chi$.*

**Theorem 3** *Let the vector field of the current probability density $\vec{J}$ admit a Helmholtz $\psi\mathbb{C}^2$ – decomposition, then the first Vlasov equation will take the form:*

$$\frac{i}{\beta} \mathrm{I} \frac{\partial \psi}{\partial t} = -\alpha\beta \left[ \vec{\sigma} \cdot \left( \hat{p} - \frac{\gamma}{2\alpha\beta} \vec{A} \right) \right]^2 \psi + \mathrm{I} U \psi, \qquad (2.3)$$

*or*

$$\frac{i}{\beta}\text{I}\frac{\partial \psi}{\partial t} = -\alpha\beta\,\text{I}\left(\hat{\text{p}} - \frac{\gamma}{2\alpha\beta}\vec{\text{A}}\right)^2 \psi + \text{I}U\psi + \frac{\gamma}{2\beta}\left(\vec{\sigma}\cdot\vec{\text{B}}\right)\psi, \qquad (2.4)$$

*where* $|\psi|^2 = f$; $\vec{\sigma} = (\sigma^1, \sigma^2, \sigma^3)$ *are the Pauli matrices;* $\sigma^0 = \text{I}$ *is an identity matrix.* $\vec{\text{B}} \stackrel{\text{def}}{=} \text{curl}_r \vec{\text{A}}$; $\hat{\text{p}} = \frac{i}{\beta}\nabla_r$, $U \in \mathbb{R}$ *is some function, and* $\alpha, \beta, \gamma$ *are some constant values.*

The proof of Theorem 3 is given in Appendix B.

When choosing constant values $\alpha, \beta, \gamma$ according to (1.6), equation (2.3)/(2.4) transforms into the well-known non-relativistic Pauli equation for a particle with spin in an external electromagnetic field [19].

$$i\hbar\,\text{I}\frac{\partial \psi}{\partial t} = \frac{1}{2m}\text{I}\left(\hat{\text{p}} - q\vec{\text{A}}\right)^2 \psi + \text{I}U\psi - \frac{q\hbar}{2m}\left(\vec{\sigma}\cdot\vec{\text{B}}\right)\psi. \qquad (2.5)$$

From a comparison of equations (1.7) and (2.5) one can see the appearance of a new term $\frac{q\hbar}{2m}\left(\vec{\sigma}\cdot\vec{\text{B}}\right)$ responsible for the spin of the particle.

Equations (1.5) and (2.3) were obtained from the first Vlasov equation using Helmholtz $\Psi\mathbb{C}-$ and $\psi\mathbb{C}^2-$ decomposition, respectively. Both equations (1.5) and (2.3) are not relativistic, since they do not have invariance under the Lorentz transformation. Note that the original Vlasov equation (i.5), ($n = 1$) is invariant. The first problem in invariance violation appears when using the Helmholtz decomposition for 3D vector field $\langle\vec{v}\rangle$ or in 3D coordinate space $\vec{\text{J}}$. Indeed, time $t$ is included in decompositions (1.1), (1.4) and (2.2) as a parameter. In the Helmholtz decomposition there are derivatives only along coordinate axes $\partial_k$, and there are no derivatives along the time axis $\partial_0$. To solve this problem, it is necessary to extend the Helmholtz decomposition to 4D space-time. On the one hand, the 4D space-time decomposition gives hope for a transition to relativism, but on the other hand, when substituting it into the Vlasov equation, a second problem arises: different orders of derivatives with respect to coordinate and time. There are two ways to solve this problem. The first is to increase the order of the derivatives in term $\partial_0 f$ from the Vlasov equation. This approach is actually implemented in the Klein-Gordon equation. The second way is to lower the order of derivatives in the first term of decomposition (1.4):

$$\text{J}^k = \frac{\hbar}{m}\Psi^*\partial^k\varphi\Psi, \qquad (2.12)$$

This expression (2.12) takes the relation (1.3) into account for the phase of wave function $\varphi$. The Helmholtz $\Psi\mathbb{C}-$decomposition (1.4) used a complex wave function $\Psi = |\Psi|\exp(i\varphi)$ characterized by modulus $|\Psi|$ and phase $\varphi$. For $\Psi \in \mathbb{C}$ the phase $\varphi$ specifies the angle of rotation in the complex plane. When transiting to the Helmholtz $\psi\mathbb{C}^2-$decomposition (2.2), spinors $\psi \in \mathbb{C}^2$ were considered instead of $\Psi \in \mathbb{C}$. Spinors $\psi$ can be acted upon by complex matrices $M$ of size $2\times 2$, that is $M\psi$. The indicated matrices can be decomposed into basis

matrices – the Pauli matrices $\sigma^\mu$, $\mu = 0...3$. The Pauli matrices correspond to the four basis elements $1, i, j, k$ in the quaternion decomposition $q = (a; \vec{u}) = a + iu_1 + ju_2 + ku_3$:

$$1 \mapsto \sigma^0 = \begin{pmatrix} 1 & 0 \\ 0 & 1 \end{pmatrix}, \; i \mapsto i\sigma^3 = i\begin{pmatrix} 1 & 0 \\ 0 & -1 \end{pmatrix}, \; j \mapsto i\sigma^2 = i\begin{pmatrix} 0 & -i \\ i & 0 \end{pmatrix}, \; k \mapsto i\sigma^1 = i\begin{pmatrix} 0 & 1 \\ 1 & 0 \end{pmatrix}. \qquad (2.13)$$

Quaternion algebra ($\mathbb{Q}$) was originally constructed by Hamilton and further developed by Clifford in the form of hypercomplex numbers [20], for example, octaves ($\mathbb{O}$) and sedenions ($\mathbb{S}$). Quaternions can be represented in the form of complex matrices $2 \times 2$ (2.13), or in the form of real matrices $4 \times 4$. Each 3D vector $\vec{r} = (x_1, x_2, x_3)$ in Euclidean space can be associated with a $2 \times 2$ matrix according to the rule

$$\vec{\sigma} \cdot \vec{r} = \sigma^k x_k = \begin{pmatrix} x_3 & x_1 - ix_2 \\ x_1 + ix_2 & -x_3 \end{pmatrix}. \qquad (2.14)$$

Representation (2.14) is used in the Pauli equation (2.3) for the momentum operator. According to spinor algebra, matrix (2.14) specifies the rotation of the spinor. Rotating a spinor from one position to another is an ambiguous operation. There are two possible rotating operations, differing only in sign. Thus, it is logical to take rotation matrices $\vec{\sigma} \cdot \vec{r}$ as phase $\varphi$. Since it is necessary to take all axes of 4D space-time $x^\mu = \begin{pmatrix} x^0 & \vec{x} \end{pmatrix}^T$ into account we define rotation matrix $\hat{\varphi}$ for bispinor $\psi \in \mathbb{C}^4$ (two directions of rotation) as follows:

$$\begin{pmatrix} \sigma^0 x_0 & \sigma^s x_s \\ \sigma^s x_s & \sigma^0 x_0 \end{pmatrix} = \begin{pmatrix} I_2 & 0 \\ 0 & -I_2 \end{pmatrix} \begin{pmatrix} \sigma^0 x_0 & \sigma^s x_s \\ -\sigma^s x_s & -\sigma^0 x_0 \end{pmatrix} = \gamma^0 \hat{\varphi}(x_\mu), \qquad (2.15)$$

$$\hat{\varphi}(x^\mu) = \begin{pmatrix} \sigma^0 x_0 & -\vec{\sigma} \cdot \vec{x} \\ \vec{\sigma} \cdot \vec{x} & -\sigma^0 x_0 \end{pmatrix}.$$

The rotation phase matrix (2.15) is defined at points of 4D space-time, that is, each point $x^\mu$ has its own rotation (phase). Note that from the Hamilton-Jacobi equation (1.8) the phase is related to the concept of action $\hbar\varphi = S$. From a geometric point of view, action $S$ corresponds to the length of the trajectory (in the case of quantum mechanics – the Feynman path integrals) in curved space with the indicatrix $\mathcal{L} = 1$, where $\mathcal{L}$ is the Lagrange [21]. Thus, for each point in 4D space-time the «trajectory length» $S$ is determined by rotation matrix (2.15).

The Vlasov equation in 4-component form will take the form:

$$\frac{\partial}{\partial ct} cf + \frac{\partial}{\partial x^k}\left[f\langle v^k \rangle\right] = 0 \; \Rightarrow \; \partial_\mu J^\mu = 0, \qquad (2.16)$$

where

$$J^\mu = \begin{pmatrix} cf & f\langle \vec{v} \rangle \end{pmatrix}^T = \begin{pmatrix} c|\psi|^2 & \vec{J} \end{pmatrix}^T, \; |\psi|^2 = \psi^\dagger \psi.$$

Let us transform representation of the current density (2.12) for bispinor $\psi \in \mathbb{C}^4$ and the phase matrix (2.15):

$$J^0 = \frac{\hbar}{m} \Psi^* \Psi \mapsto c\psi^\dagger \psi, \tag{2.17}$$

$$J^k = \frac{\hbar}{m} \Psi^* \partial^k \hat{\varphi} \Psi \mapsto c\psi^\dagger \gamma^0 \partial^k \hat{\varphi} \psi,$$

where, in contrast to (1.4), density $f = |\psi|^2$ is a dimensionless value. Let us calculate expression $\partial^k \hat{\varphi}$ from (2.17). The following expression will be obtained:

$$\partial^k \hat{\varphi} = \begin{pmatrix} 0 & \sigma^k \\ -\sigma^k & 0 \end{pmatrix}, \quad \partial^0 \hat{\varphi} = \begin{pmatrix} I_2 & 0 \\ 0 & -I_2 \end{pmatrix},$$

$$\partial^\mu \hat{\varphi} = \gamma^\mu, \tag{2.18}$$

where $\gamma^\mu$ are the Dirac matrices. From a physical point of view, expression (2.18) can have the following interpretation. The partial derivative of the phase (action) with respect to time $\partial^0 \hat{\varphi}$ in accordance with the Hamilton-Jacobi equation (1.8) determines the energy of the system. From expression (2.18) it follows that there are positive and negative values of the system energy (a particle and an antiparticle). In accordance with the Helmholtz decomposition (1.1) the derivative with respect to coordinate space from the phase (action) $\partial^k \hat{\varphi}$ is associated with the potential component of the momentum $\langle \vec{p} \rangle = \hbar \nabla_r \varphi$. From expression (2.18) it follows that the momentum is related to the Pauli matrices $\sigma^k$, which corresponds to the Pauli equation (2.3) $\vec{\sigma} \cdot \hat{p}$.

Taking expression (2.18) into account the representation (2.17) for current density $J^\mu$ will take the form:

$$J^\mu = c\bar{\psi}\gamma^\mu \psi, \tag{2.19}$$

where $\bar{\psi} \stackrel{\text{def}}{=} \psi^\dagger \gamma^0$ is a dual wave function and $\gamma^0 \gamma^0 = I$ is an identity matrix.

**Theorem 4** *Let the probability current density $J^\mu$ admit the (2.19) representation, then the Vlasov equation (2.16) will take the form:*

$$i\hbar c \partial_0 \psi = \gamma^0 \left[ c\gamma^k \left( \hat{p}_k - q A_k \right) + mc^2 I \right] \psi + q\varphi\psi, \tag{2.20}$$

*or*

$$\left[ \gamma^\mu \left( i\hbar \partial_\mu - qA_\mu \right) - mc I \right] \psi = 0,$$

*where $m, q, \hbar$ are constant values, $\hat{p}_k = -i\hbar \partial_k$, and $A^\mu = (\varphi/c, \vec{A})$ is some four-vector field.*

The proof of Theorem 4 is given in Appendix B.

Equation (2.20) is known as the relativistic Dirac equation for a particle (and an antiparticle) in an electromagnetic field [22].

**Conclusions**

Quantum mechanics, like no other branch of physics, contains a large number of postulates, correspondence rules and principles that were introduced in a phenomenological manner. This construction of quantum theory is caused by the strong «dissimilarity» of the behavior of the micro and macro worlds. Historical attempts to «understand» the quantum world are well illustrated by the examples of dialogues between Einstein and Bohr, the EPR paradox, the theory with hidden parameters, and many others. The fatigue from fruitless attempts to understand the micro world nature called into existence Feynman's famous phrase *«I think I can safely say: no one understands quantum mechanics»*.

Over the last century, quantum theory has made significant progress, but despite this, the physical nature of the micro world remains unclear. From the position of theoretical physics, it is important not only to know the rules, laws and equations, but also to build a mathematical model, to find some first principle which can explain the behavior of the system. For example, Newton's second law was obtained phenomenologically, but the principle of least action (PLA) made it possible to interpret it as the shortest distance between two positions of a system in some curved space, where the curvature is caused by an external influence. Feynman's path integral gave us a classical interpretation of the quantum dynamics of a system from the perspective of PLA.

The beauty of a theory is determined by the number of postulates, rules, and axioms contained in it. The more rules, the more hidden the principle of the physical system is. In this paper, an attempt is made to outline a method for constructing some equations of classical and quantum mechanics based on one single first principle – the probability conservation law. The central object of this approach is the function of distribution or density of probabilities that satisfies the probability conservation law. An expanded version of the theory presented is given in [23-26], herein its application to the description of quantum systems and their relation with classical systems is shown.

**Appendix A**
*Proof of Theorem 1*

Let us make intermediate transformations

$$\Psi^* \frac{\partial \Psi}{\partial t} + \Psi \frac{\partial \Psi^*}{\partial t} + i\alpha \, \text{div}_r \left[ \Psi^* \nabla \Psi - \Psi \nabla \Psi^* - i\frac{\gamma}{\alpha} \Psi \Psi^* \vec{A}_\Psi \right] = 0. \qquad (A.1)$$

taking into account that

$$\text{div}_r \left[ \Psi^* \nabla_r \Psi - \Psi \nabla_r \Psi^* \right] = \Psi^* \Delta_r \Psi - \Psi \Delta_r \Psi^*,$$

$$\text{div}_r \left[ \Psi \Psi^* \vec{A}_\Psi \right] = \Psi^* \vec{A}_\Psi \cdot \nabla_r \Psi + \Psi \vec{A}_\Psi \cdot \nabla_r \Psi^* + \Psi \Psi^* \chi, \qquad (A.2)$$

we obtain

$$\Psi^* \left[ \frac{\partial}{\partial t} + i\alpha\Delta + \gamma \vec{A}_\Psi \cdot \nabla + \frac{\gamma}{2}\chi \right] \Psi + \Psi \left[ \frac{\partial}{\partial t} - i\alpha\Delta + \gamma \vec{A}_\Psi \cdot \nabla + \frac{\gamma}{2}\chi \right] \Psi^* = 0, \qquad (A.3)$$

or

$$\Psi^* \left[ \frac{1}{\beta} \frac{\partial}{\partial t} - i\alpha\beta \left( \hat{p}^2 - \frac{\gamma}{\alpha\beta} \vec{A}_\Psi \cdot \hat{p} \right) + \frac{\gamma}{2\beta} \chi \right] \Psi +$$
$$+ \Psi \left[ \frac{1}{\beta} \frac{\partial}{\partial t} + i\alpha\beta \left( \hat{p}^{*2} - \frac{\gamma}{\alpha\beta} \vec{A}_\Psi \cdot \hat{\bar{p}} \right) + \frac{\gamma}{2\beta} \chi \right] \Psi^* = 0. \qquad (A.4)$$

We take into account the ratios:

$$\hat{p}^2 - \frac{\gamma}{\alpha\beta}\vec{A}_\Psi \cdot \hat{p} = \left(\hat{p} - \frac{\gamma}{2\alpha\beta}\vec{A}_\Psi\right)^2 - \frac{\gamma^2}{4\alpha^2\beta^2}\left|\vec{A}_\Psi\right|^2 - i\frac{\gamma}{2\alpha\beta^2}\chi. \qquad (A.5)$$

Substituting (A.5) into equation (A.4), we obtain

$$\Psi^*\left[\frac{1}{\beta}\frac{\partial}{\partial t} - i\alpha\beta\left(\hat{p} - \frac{\gamma}{2\alpha\beta}\vec{A}_\Psi\right)^2\right]\Psi + \Psi\left[\frac{1}{\beta}\frac{\partial}{\partial t} + i\alpha\beta\left(\hat{p}^* - \frac{\gamma}{2\alpha\beta}\vec{A}_\Psi\right)^2\right]\Psi^* = 0, \qquad (A.6)$$

For convenience, we introduce operator $L$:

$$L = \frac{1}{\beta}\frac{\partial}{\partial t} - i\alpha\beta\left(\hat{p} - \frac{\gamma}{2\alpha\beta}\vec{A}_\Psi\right)^2. \qquad (A.7)$$

Let us rewrite equation (A.6) using the operator (A.7):

$$\Psi^* L\Psi + \Psi L^*\Psi^* = 0 \Rightarrow \Lambda + \Lambda^* = 0, \ \Lambda = \Psi^* L\Psi. \qquad (A.8)$$

from here

$$\mathrm{Re}\,\Lambda = 0 \Rightarrow \Lambda = iu,\ u \in \mathbb{R}. \qquad (A.9)$$

From expressions (A.8) and (A.9), we obtain the equation

$$\Psi^* L\Psi = iu \Rightarrow L\Psi = i\frac{u}{\Psi^*} = i\frac{u}{\Psi^*}\frac{\Psi}{\Psi} = i\frac{u}{|\Psi|^2}\Psi = -iU\Psi,\ U \in \mathbb{R},$$
$$L\Psi = -iU\Psi. \qquad (A.10)$$

Substituting the form of operator (A.7) into (A.10), we obtain the equation:

$$\frac{i}{\beta}\frac{\partial \Psi}{\partial t} = -\alpha\beta\left(\hat{p} - \frac{\gamma}{2\alpha\beta}\vec{A}_\Psi\right)^2\Psi + U\Psi. \qquad (A.11)$$

Theorem 1 is proved

*Proof of Theorem 2*

Using (A.11) and (A.5), we obtain an expression for function $U$:

$$U\Psi = \frac{i}{\beta}\frac{\partial \Psi}{\partial t} - \frac{\alpha}{\beta}\Delta_r\Psi + i\frac{\gamma}{\beta}\vec{A}\cdot\nabla_r\Psi + \frac{\gamma^2}{4\alpha\beta}\left|\vec{A}_\Psi\right|^2\Psi + i\frac{\gamma}{2\beta}\chi\Psi. \qquad (A.12)$$

Let us calculate $\Delta_r\Psi$

$$\Delta_r\Psi = e^{i\varphi}\left[\Delta_r|\Psi| - |\Psi|\left|\nabla_r\varphi\right|^2 + i\left(2\nabla_r\varphi\cdot\nabla_r|\Psi| + |\Psi|\Delta_r\varphi\right)\right]. \qquad (A.13)$$

We substitute (A.13) into (A.12), we obtain

$$U = \frac{i}{\beta|\Psi|}\frac{\partial|\Psi|}{\partial t} - i\frac{\alpha}{\beta}\left(\frac{2}{|\Psi|}\nabla_r\varphi\cdot\nabla_r|\Psi| + \Delta_r\varphi\right) + i\frac{\gamma}{|\Psi|\beta}\vec{A}_\Psi\cdot\nabla_r|\Psi| +$$

$$+ i\frac{\gamma}{2\beta}\chi - \frac{1}{\beta}\frac{\partial\varphi}{\partial t} - \frac{\alpha}{\beta}\frac{\Delta_r|\Psi|}{|\Psi|} + \frac{\alpha}{\beta}|\nabla_r\varphi|^2 - \frac{\gamma}{\beta}\vec{A}_\Psi\cdot\nabla_r\varphi + \frac{\gamma^2}{4\alpha\beta}|\vec{A}_\Psi|^2. \tag{A.14}$$

From expression (A.10) it follows that the imaginary part (A.14) must be equal to zero. Indeed, expression (A.14) takes the form:

$$\begin{cases} \dfrac{\partial f}{\partial t} + \left(-\alpha\nabla_r\Phi + \gamma\vec{A}_\Psi\right)\cdot\nabla_r f + f\left(-\alpha\Delta_r\Phi + \gamma\,\mathrm{div}_r\,\vec{A}_\Psi\right) = 0, \\ -\dfrac{1}{\beta}\dfrac{\partial\varphi}{\partial t} - Q + \dfrac{1}{4\alpha\beta}\left(|\alpha\nabla_r\Phi|^2 - 2\alpha\nabla_r\Phi\cdot\gamma\vec{A}_\Psi + |\gamma\vec{A}_\Psi|^2\right) - U = 0. \end{cases}$$

from here

$$\begin{cases} \dfrac{\partial f}{\partial t} + \langle\vec{v}\rangle\cdot\nabla_r f + f\,\mathrm{div}_r\,\langle\vec{v}\rangle = 0, \\ -\dfrac{1}{\beta}\dfrac{\partial\varphi}{\partial t} = -\dfrac{1}{4\alpha\beta}|\langle\vec{v}\rangle|^2 + U + Q. \end{cases} \tag{A.15}$$

The first equation in (A.15) coincides with the first Vlasov equation, and the second leads to the Hamilton-Jacobi equation.

$$-\frac{1}{\beta}\frac{\partial\varphi}{\partial t} = -\frac{1}{4\alpha\beta}|\langle\vec{v}\rangle|^2 + V = H, \tag{A.16}$$

$$V = U + Q, \quad Q = \frac{\alpha}{\beta}\frac{\Delta_r|\Psi|}{|\Psi|}. \tag{A.17}$$

We obtain the equation of motion from equation (A.16):

$$\frac{1}{2\alpha\beta}\frac{\partial}{\partial t}(-\alpha\nabla_r\Phi) = \frac{1}{2\alpha\beta}\frac{\partial}{\partial t}\langle\vec{v}\rangle - \frac{\gamma}{2\alpha\beta}\frac{\partial}{\partial t}\vec{A}_\Psi = -\frac{1}{4\alpha\beta}\nabla_r|\langle\vec{v}\rangle|^2 + \nabla_r V, \tag{A.18}$$

taking into account that $(\langle\vec{v}\rangle\cdot\nabla_r)\langle\vec{v}\rangle = \dfrac{1}{2}\nabla_r|\langle\vec{v}\rangle|^2 - \gamma\langle\vec{v}\rangle\times\mathrm{curl}_r\,\vec{A}_\Psi$, then (A.18) takes the form:

$$\frac{\partial}{\partial t}\langle\vec{v}\rangle + (\langle\vec{v}\rangle\cdot\nabla_r)\langle\vec{v}\rangle = \gamma\frac{\partial\vec{A}_\Psi}{\partial t} - \gamma\langle\vec{v}\rangle\times\vec{B}_\Psi + 2\alpha\beta\nabla_r V =$$

$$= -\gamma\left(-\frac{\partial\vec{A}_\Psi}{\partial t} - \frac{2\alpha\beta}{\gamma}\nabla_r V + \langle\vec{v}\rangle\times\vec{B}_\Psi\right),$$

$$\frac{d}{dt}\langle\vec{v}\rangle = -\gamma\left(\vec{E}_\Psi + \langle\vec{v}\rangle\times\vec{B}_\Psi\right), \quad \vec{E}_\Psi = -\frac{\partial\vec{A}_\Psi}{\partial t} - \frac{2\alpha\beta}{\gamma}\nabla_r V.$$

Theorem 2 is proved.

***Proof of Lemma 1***

From the condition (1.17) and the first Vlasov equation it follows that:

$$\frac{\partial}{\partial t}\operatorname{div}_r \vec{D}_\Psi + \operatorname{div}_r\left(\rho_\Psi \langle \vec{v}\rangle\right) = 0 \Rightarrow \operatorname{div}_r\left(\frac{\partial}{\partial t}\vec{D}_\Psi + \vec{J}_\Psi\right) = 0,$$

$$\frac{\partial}{\partial t}\vec{D}_\Psi + \vec{J}_\Psi = \operatorname{curl}_r \vec{H}_\Psi, \tag{A.19}$$

where $\vec{H}_\Psi$ is some field, since $\operatorname{div}_r \operatorname{curl}_r \vec{H}_\Psi = 0$. The definition of vortex field $\vec{B}_\Psi$ (1.11) leads to the equation

$$\operatorname{div}_r \vec{B}_\Psi = \operatorname{div}_r \operatorname{curl}_r \vec{A}_\Psi = 0. \tag{A.20}$$

Calculating operator $\operatorname{curl}_r$ from expression (1.11) $\vec{E}_\Psi$ gives (1.17):

$$\operatorname{curl}_r \vec{E}_\Psi = -\frac{\partial}{\partial t}\operatorname{curl}_r \vec{A}_\Psi = -\frac{\partial}{\partial t}\vec{B}_\Psi. \tag{A.21}$$

Resulting equations (A.19)-(A.21) prove the validity of equations (1.16)-(1.17). We obtain equations (1.18)-(1.20) under the condition of Lorentz $\Psi$–gauge. Calculating the operator $\operatorname{div}_r$ from expression (1.11) and taking into account the gauge condition (1.12), we obtain

$$\operatorname{div}_r \vec{E}_\Psi = \frac{\rho_\Psi}{\varepsilon_0} = -\frac{\partial}{\partial t}\operatorname{div}_r \vec{A}_\Psi - \frac{2\alpha\beta}{\gamma}\Delta_r V = \frac{2\alpha\beta}{\gamma}\left(\frac{1}{c^2}\frac{\partial^2 V}{\partial t^2} - \Delta_r V\right) = \frac{1}{q}\square V. \tag{A.22}$$

Using representations (1.11) in equation (1.17), we obtain the equation:

$$-\varepsilon_0 \frac{\partial^2}{\partial t^2}\vec{A}_\Psi - \varepsilon_0 \nabla_r \frac{\partial}{\partial t}V + \vec{J}_\Psi = \frac{1}{\mu_0}\operatorname{curl}_r \operatorname{curl}_r \vec{A}_\Psi,$$

$$-\frac{1}{c^2}\frac{\partial^2}{\partial t^2}\vec{A}_\Psi - \nabla_r\left(\frac{1}{c^2}\frac{\partial}{\partial t}V + \operatorname{div}_r \vec{A}_\Psi\right) + \Delta_r \vec{A}_\Psi = -\mu_0 \vec{J}_\Psi,$$

$$\left(\frac{1}{c^2}\frac{\partial^2}{\partial t^2} - \Delta_r\right)\vec{A}_\Psi = \mu_0 \vec{J}_\Psi. \tag{A.23}$$

Expressions (A.22) and (A.23) are consistent with equations (1.18)-(1.20). Lemma 1 is completely proved.

**Appendix B**
*Proof of Theorem 3*

Substituting the Helmholtz $\psi\mathbb{C}^2$–decomposition (3.2) into the first Vlasov equation, we obtain

$$\psi^\dagger \frac{\partial \psi}{\partial t} + \frac{\partial \psi^\dagger}{\partial t}\psi + i\alpha \operatorname{div}_r\left[\psi^\dagger \mathrm{I}\nabla_r \psi - \psi^T \mathrm{I}\nabla_r \psi^* - i\frac{\gamma}{\alpha}\psi^\dagger \vec{A}\psi\right] = 0, \qquad \mathrm{I}\nabla_r \stackrel{\text{det}}{=}\begin{pmatrix}\nabla_r & 0 \\ 0 & \nabla_r\end{pmatrix}. \tag{B.1}$$

Let us perform intermediate transformations:

$$\psi^\dagger \frac{\partial \psi}{\partial t} + \frac{\partial \psi^\dagger}{\partial t}\psi = \psi_1^* \partial_t \psi_1 + \psi_2^* \partial_t \psi_2 + \psi_1 \partial_t \psi_1^* + \psi_2 \partial_t \psi_2^* =$$

$$= \begin{pmatrix} \psi_1^* & \psi_2^* \end{pmatrix} \begin{pmatrix} \partial_t & 0 \\ 0 & \partial_t \end{pmatrix} \begin{pmatrix} \psi_1 \\ \psi_2 \end{pmatrix} + \begin{pmatrix} \psi_1 & \psi_2 \end{pmatrix} \begin{pmatrix} \partial_t & 0 \\ 0 & \partial_t \end{pmatrix} \begin{pmatrix} \psi_1^* \\ \psi_2^* \end{pmatrix},$$

$$\psi^\dagger \frac{\partial \psi}{\partial t} + \frac{\partial \psi^\dagger}{\partial t}\psi = \psi^\dagger \mathrm{I} \partial_t \psi + \psi^T \mathrm{I} \partial_t \psi^*, \tag{B.2}$$

$$\psi^\dagger \mathrm{I} \nabla_r \psi - \psi^T \mathrm{I} \nabla_r \psi^* = \psi_1^* \nabla_r \psi_1 + \psi_2^* \nabla_r \psi_2 - \psi_1 \nabla_r \psi_1^* - \psi_2 \nabla_r \psi_2^*, \tag{B.3}$$

$$\mathrm{div}_r \left( \psi^\dagger \mathrm{I} \nabla_r \psi - \psi^T \mathrm{I} \nabla_r \psi^* \right) = \psi^\dagger \mathrm{I} \Delta_r \psi - \psi^T \mathrm{I} \Delta_r \psi^*, \tag{B.4}$$

$$\mathrm{div}_r \left( \psi^\dagger \vec{\mathrm{A}} \psi \right) = \psi^\dagger \chi \psi + \psi^T \mathrm{I} \vec{\mathrm{A}} \cdot \nabla_r \psi^* + \psi^\dagger \mathrm{I} \vec{\mathrm{A}} \cdot \nabla_r \psi. \tag{B.5}$$

Substituting expressions (B.2)-(B.4) into equation (B.1), we obtain:

$$\psi^\dagger \mathrm{I} \left[ \frac{1}{\beta} \frac{\partial}{\partial t} - i\alpha\beta \left( \hat{\mathrm{p}}^2 - \frac{\gamma}{\alpha\beta} \vec{\mathrm{A}} \cdot \hat{\mathrm{p}} \right) + \frac{\gamma}{2\beta} \chi \right] \psi + \psi^T \mathrm{I} \left[ \frac{1}{\beta} \frac{\partial}{\partial t} + i\alpha\beta \left( \hat{\mathrm{p}}^{*2} - \frac{\gamma}{\alpha\beta} \vec{\mathrm{A}} \cdot \hat{\mathrm{p}}^* \right) + \frac{\gamma}{2\beta} \chi \right] \psi^* = 0, \tag{B.6}$$

Let us transform the terms in equation (B.5):

$$\mathrm{I} \left( \hat{\mathrm{p}}^2 - \frac{\gamma}{\alpha\beta} \vec{\mathrm{A}} \cdot \hat{\mathrm{p}} \right) = \mathrm{I} \left( \hat{\mathrm{p}} - \frac{\gamma}{2\alpha\beta} \vec{\mathrm{A}} \right)^2 - \mathrm{I} \frac{\gamma^2}{4\alpha^2 \beta^2} |\vec{\mathrm{A}}|^2 - i\mathrm{I} \frac{\gamma}{2\alpha\beta^2} \chi =$$

$$= \left[ \vec{\sigma} \cdot \left( \hat{\mathrm{p}} - \frac{\gamma}{2\alpha\beta} \vec{\mathrm{A}} \right) \right]^2 + \frac{\gamma}{2\alpha\beta^2} \left( \vec{\sigma} \cdot \mathrm{curl}_r \vec{\mathrm{A}} - \mathrm{I} i\chi \right) - \mathrm{I} \frac{\gamma^2}{4\alpha^2 \beta^2} |\vec{\mathrm{A}}|^2, \tag{B.7}$$

where, in accordance with expression $(\vec{\sigma} \cdot \vec{a})(\vec{\sigma} \cdot \vec{b}) = \mathrm{I}(\vec{a} \cdot \vec{b}) + i\vec{\sigma} \cdot (\vec{a} \times \vec{b})$, it is taken into account that

$$\left[ \vec{\sigma} \cdot \left( \hat{\mathrm{p}} - \frac{\gamma}{2\alpha\beta} \vec{\mathrm{A}} \right) \right]^2 = \mathrm{I} \left( \hat{\mathrm{p}} - \frac{\gamma}{2\alpha\beta} \vec{\mathrm{A}} \right)^2 - \frac{\gamma}{2\alpha\beta^2} \vec{\sigma} \cdot \mathrm{curl}_r \vec{\mathrm{A}}. \tag{B.8}$$

Substituting (B.7) into equation (B.6), we obtain

$$\psi^\dagger \left( \frac{1}{\beta} \mathrm{I} \frac{\partial}{\partial t} - i\alpha\beta \left[ \vec{\sigma} \cdot \left( \hat{\mathrm{p}} - \frac{\gamma}{2\alpha\beta} \vec{\mathrm{A}} \right) \right]^2 + i\mathrm{I} \frac{\gamma^2}{4\alpha\beta} |\vec{\mathrm{A}}|^2 \right) \psi + \tag{B.9}$$

$$+\psi^T \left( \frac{1}{\beta} \mathrm{I} \frac{\partial}{\partial t} + i\alpha\beta \left[ \vec{\sigma}^* \cdot \left( \hat{\mathrm{p}}^* - \frac{\gamma}{2\alpha\beta} \vec{\mathrm{A}} \right) \right]^2 - i\mathrm{I} \frac{\gamma^2}{4\alpha\beta} |\vec{\mathrm{A}}|^2 \right) \psi^* + i\frac{\gamma}{2\beta} \left[ \psi^T \left( \vec{\sigma}^* \cdot \vec{\mathrm{B}} \right) \psi^* - \psi^\dagger \left( \vec{\sigma} \cdot \vec{\mathrm{B}} \right) \psi \right] = 0.$$

Let us simplify the third term in equation (B.9):

$$\psi^T \left( \vec{\sigma}^* \cdot \vec{\mathrm{B}} \right) \psi^* - \psi^\dagger \left( \vec{\sigma} \cdot \vec{\mathrm{B}} \right) \psi = 0. \tag{B.10}$$

Taking into account expression (B.10), equation (B.9) will take the form

$$\psi^{\dagger}\left(\frac{1}{\beta}\mathrm{I}\frac{\partial}{\partial t}-i\alpha\beta\left[\vec{\sigma}\cdot\left(\hat{\mathrm{p}}-\frac{\gamma}{2\alpha\beta}\vec{\mathrm{A}}\right)\right]^{2}\right)\psi+i\frac{\gamma^{2}}{4\alpha\beta}\left|\vec{\mathrm{A}}\right|^{2}\left(\psi^{\dagger}\mathrm{I}\psi-\psi^{T}\mathrm{I}\psi^{*}\right)+$$

$$+\psi^{T}\left(\frac{1}{\beta}\mathrm{I}\frac{\partial}{\partial t}+i\alpha\beta\left[\vec{\sigma}^{*}\cdot\left(\hat{\mathrm{p}}^{*}-\frac{\gamma}{2\alpha\beta}\vec{\mathrm{A}}\right)\right]^{2}\right)\psi^{*}=0. \qquad (B.11)$$

The second term in equation (B.11) is equal to zero, indeed

$$\psi^{\dagger}\mathrm{I}\psi-\psi^{T}\mathrm{I}\psi^{*}=\psi^{\dagger}\psi-\psi^{T}\psi^{*}=\left|\psi\right|^{2}-\left|\psi\right|^{2}=0,$$

from here

$$\psi^{\dagger}\left(\frac{1}{\beta}\mathrm{I}\frac{\partial}{\partial t}-i\alpha\beta\left[\vec{\sigma}\cdot\left(\hat{\mathrm{p}}-\frac{\gamma}{2\alpha\beta}\vec{\mathrm{A}}\right)\right]^{2}\right)\psi+\psi^{T}\left(\frac{1}{\beta}\mathrm{I}\frac{\partial}{\partial t}+i\alpha\beta\left[\vec{\sigma}^{*}\cdot\left(\hat{\mathrm{p}}^{*}-\frac{\gamma}{2\alpha\beta}\vec{\mathrm{A}}\right)\right]^{2}\right)\psi^{*}=0. \qquad (B.12)$$

For convenience, we introduce operator $L$:

$$L \stackrel{\text{det}}{=} \frac{1}{\beta}\mathrm{I}\frac{\partial}{\partial t}-i\alpha\beta\left[\vec{\sigma}\cdot\left(\hat{\mathrm{p}}-\frac{\gamma}{2\alpha\beta}\vec{\mathrm{A}}\right)\right]^{2}, \ L^{*}=\frac{1}{\beta}\mathrm{I}\frac{\partial}{\partial t}+i\alpha\beta\left[\vec{\sigma}^{*}\cdot\left(\hat{\mathrm{p}}^{*}-\frac{\gamma}{2\alpha\beta}\vec{\mathrm{A}}\right)\right]^{2}. \qquad (B.13)$$

Let us rewrite equation (B.12) using operator (B.13)

$$\psi^{\dagger}L\psi+\psi^{T}L^{*}\psi^{*}=0 \Rightarrow \Lambda+\Lambda^{*}=0, \ \Lambda\stackrel{\text{det}}{=}\psi^{\dagger}L\psi. \qquad (B.14)$$

Since $\Lambda\in\mathbb{C}^{[2\times 2]}$, from here $\Lambda+\Lambda^{*}=2\operatorname{Re}\Lambda$

$$\operatorname{Re}\Lambda=0 \Rightarrow \Lambda=iu, \ u\in\mathbb{R}. \qquad (B.15)$$

From expressions (B.14) and (B.15) we obtain the equation

$$\psi^{\dagger}L\psi=-i\left|\psi\right|^{2}U=-i\psi^{\dagger}\mathrm{I}U\psi \Rightarrow \psi^{\dagger}\cdot\left(L\psi+i\mathrm{I}U\psi\right)=0, \ U\in\mathbb{R},$$
$$L\psi=-i\mathrm{I}U\psi. \qquad (B.16)$$

Substituting the form of operator (B.13) into (B.16) we arrive at the equation:

$$\frac{i}{\beta}\mathrm{I}\frac{\partial\psi}{\partial t}=-\alpha\beta\left[\vec{\sigma}\cdot\left(\hat{\mathrm{p}}-\frac{\gamma}{2\alpha\beta}\vec{\mathrm{A}}\right)\right]^{2}\psi+\mathrm{I}U\psi. \qquad (B.17)$$

Theorem 3 is proved.

***Proof of Theorem 4***

Bispinor $\psi$ can be represented in terms of two spinors $\vartheta$ and $\eta$:

$$\psi=\left(\vartheta \quad \eta\right)^{T}=\left(\psi_{1} \quad \psi_{2} \quad \psi_{3} \quad \psi_{4}\right)^{T}, \ \vartheta=\left(\psi_{1} \quad \psi_{2}\right)^{T}, \ \eta=\left(\psi_{3} \quad \psi_{4}\right)^{T},$$

$$|\psi|^2 = \psi^\dagger \psi = |\vartheta|^2 + |\eta|^2, \tag{B.21}$$

$$\bar{\psi} = \psi^\dagger \gamma^0 = \begin{pmatrix} \vartheta^\dagger & \eta^\dagger \end{pmatrix} \begin{pmatrix} I_2 & 0 \\ 0 & -I_2 \end{pmatrix} = \begin{pmatrix} \vartheta^\dagger & -\eta^\dagger \end{pmatrix} = \begin{pmatrix} \psi_1^* & \psi_2^* & -\psi_3^* & -\psi_4^* \end{pmatrix}.$$

Let us write expressions for density of current $J^\mu$:

$$J^0 = c\bar{\psi}\gamma^0\psi = c\psi^\dagger \gamma^0\gamma^0\psi = c\psi^\dagger\psi = c\left(\psi_1^*\psi_1 + \psi_2^*\psi_2 + \psi_3^*\psi_3 + \psi_4^*\psi_4\right), \tag{B.22}$$

$$J^k = c\psi^\dagger \gamma^0\gamma^k\psi = c\begin{pmatrix} \vartheta^\dagger & -\eta^\dagger \end{pmatrix}\begin{pmatrix} 0 & \sigma^k \\ -\sigma^k & 0 \end{pmatrix}\begin{pmatrix} \vartheta \\ \eta \end{pmatrix} = c\begin{pmatrix} \vartheta^\dagger & -\eta^\dagger \end{pmatrix}\begin{pmatrix} \sigma^k\eta \\ -\sigma^k\vartheta \end{pmatrix},$$

$$J^k = c\left(\vartheta^\dagger \sigma^k \eta + \eta^\dagger \sigma^k \vartheta\right). \tag{B.23}$$

Hence,

$$J^1 = c\left(\vartheta^\dagger \sigma^1 \eta + \eta^\dagger \sigma^1 \vartheta\right) = c\begin{pmatrix} \psi_1^* & \psi_2^* \end{pmatrix}\begin{pmatrix} 0 & 1 \\ 1 & 0 \end{pmatrix}\begin{pmatrix} \psi_3 \\ \psi_4 \end{pmatrix} + c\begin{pmatrix} \psi_3^* & \psi_4^* \end{pmatrix}\begin{pmatrix} 0 & 1 \\ 1 & 0 \end{pmatrix}\begin{pmatrix} \psi_1 \\ \psi_2 \end{pmatrix} =$$

$$= c\left(\psi_1^*\psi_4 + \psi_4^*\psi_1 + \psi_2^*\psi_3 + \psi_3^*\psi_2\right) = 2c\,\mathrm{Re}\left(\psi_1^*\psi_4 + \psi_2^*\psi_3\right),$$

$$J^2 = c\left(\vartheta^\dagger \sigma^2 \eta + \eta^\dagger \sigma^2 \vartheta\right) = c\begin{pmatrix} \psi_1^* & \psi_2^* \end{pmatrix}\begin{pmatrix} 0 & -i \\ i & 0 \end{pmatrix}\begin{pmatrix} \psi_3 \\ \psi_4 \end{pmatrix} + c\begin{pmatrix} \psi_3^* & \psi_4^* \end{pmatrix}\begin{pmatrix} 0 & -i \\ i & 0 \end{pmatrix}\begin{pmatrix} \psi_1 \\ \psi_2 \end{pmatrix} = \tag{B.24}$$

$$= ic\left(\psi_2^*\psi_3 - \psi_1^*\psi_4 + \psi_4^*\psi_1 - \psi_3^*\psi_2\right) = -2c\,\mathrm{Im}\left(\psi_2^*\psi_3 - \psi_1^*\psi_4\right),$$

$$J^3 = c\left(\vartheta^\dagger \sigma^3 \eta + \eta^\dagger \sigma^3 \vartheta\right) = c\begin{pmatrix} \psi_1^* & \psi_2^* \end{pmatrix}\begin{pmatrix} 1 & 0 \\ 0 & -1 \end{pmatrix}\begin{pmatrix} \psi_3 \\ \psi_4 \end{pmatrix} + c\begin{pmatrix} \psi_3^* & \psi_4^* \end{pmatrix}\begin{pmatrix} 1 & 0 \\ 0 & -1 \end{pmatrix}\begin{pmatrix} \psi_1 \\ \psi_2 \end{pmatrix} =$$

$$= c\left(\psi_1^*\psi_3 - \psi_4^*\psi_2 + \psi_3^*\psi_1 - \psi_2^*\psi_4\right) = 2c\,\mathrm{Re}\left(\psi_1^*\psi_3 - \psi_4^*\psi_2\right).$$

Let us substitute the expressions for bispinors (B.22) and currents (B.24) into the first Vlasov equation (2.16)

$$\begin{aligned}
&\psi_1\left(\partial_0\psi_1^* + \partial_x\psi_4^* + i\partial_y\psi_4^* + \partial_z\psi_3^*\right) + \psi_1^*\left(\partial_0\psi_1 + \partial_x\psi_4 - i\partial_y\psi_4 + \partial_z\psi_3\right) \\
&+\psi_2\left(\partial_0\psi_2^* + \partial_x\psi_3^* - i\partial_y\psi_3^* - \partial_z\psi_4^*\right) + \psi_2^*\left(\partial_0\psi_2 + \partial_x\psi_3 + i\partial_y\psi_3 - \partial_z\psi_4\right) \\
&+\psi_3\left(\partial_0\psi_3^* + \partial_x\psi_2^* + i\partial_y\psi_2^* + \partial_z\psi_1^*\right) + \psi_3^*\left(\partial_0\psi_3 + \partial_x\psi_2 - i\partial_y\psi_2 + \partial_z\psi_1\right) \\
&+\psi_4\left(\partial_0\psi_4^* + \partial_x\psi_1^* - i\partial_y\psi_1^* - \partial_z\psi_2^*\right) + \psi_4^*\left(\partial_0\psi_4 + \partial_x\psi_1 + i\partial_y\psi_1 - \partial_z\psi_2\right) = 0.
\end{aligned} \tag{B.25}$$

We take into account that

$$J^k A_k - J^k A_k = 0,$$

$$\frac{1}{c}J^k A_k = \psi_1^*\left(A_x\psi_4 - iA_y\psi_4 + A_z\psi_3\right) + \psi_2^*\left(A_x\psi_3 + iA_y\psi_3 - A_z\psi_4\right) + \tag{B.26}$$

$$+ \psi_3^*\left(A_x\psi_2 - iA_y\psi_2 + A_z\psi_1\right) + \psi_4^*\left(A_x\psi_1 + iA_y\psi_1 - A_z\psi_2\right).$$

where $A_k$ is some vector field. Substituting (B.26) into equation (B.25), we obtain

$$\psi_1\left(\partial_0\psi_1^* + \partial_x\psi_4^* + i\partial_y\psi_4^* + \partial_z\psi_3^*\right) + \psi_1^*\left(\partial_0\psi_1 + \partial_x\psi_4 - i\partial_y\psi_4 + \partial_z\psi_3 - i\kappa A_x\psi_4 - \kappa A_y\psi_4 - i\kappa A_z\psi_3\right)$$
$$+\psi_2\left(\partial_0\psi_2^* + \partial_x\psi_3^* - i\partial_y\psi_3^* - \partial_z\psi_4^*\right) + \psi_2^*\left(\partial_0\psi_2 + \partial_x\psi_3 + i\partial_y\psi_3 - \partial_z\psi_4 - i\kappa A_x\psi_3 + \kappa A_y\psi_3 + i\kappa A_z\psi_4\right)$$
$$+\psi_3\left(\partial_0\psi_3^* + \partial_x\psi_2^* + i\partial_y\psi_2^* + \partial_z\psi_1^*\right) + \psi_3^*\left(\partial_0\psi_3 + \partial_x\psi_2 - i\partial_y\psi_2 + \partial_z\psi_1 - i\kappa A_x\psi_2 - \kappa A_y\psi_2 - i\kappa A_z\psi_1\right)$$
$$+\psi_4\left(\partial_0\psi_4^* + \partial_x\psi_1^* - i\partial_y\psi_1^* - \partial_z\psi_2^*\right) + \psi_4^*\left(\partial_0\psi_4 + \partial_x\psi_1 + i\partial_y\psi_1 - \partial_z\psi_2 - i\kappa A_x\psi_1 + \kappa A_y\psi_1 + i\kappa A_z\psi_2\right)$$
$$+i\kappa\psi_1^*\left(A_x\psi_4 - iA_y\psi_4 + A_z\psi_3\right) + i\kappa\psi_2^*\left(A_x\psi_3 + iA_y\psi_3 - A_z\psi_4\right) + i\kappa\psi_3^*\left(A_x\psi_2 - iA_y\psi_2 + A_z\psi_1\right) +$$
$$+i\kappa\psi_4^*\left(A_x\psi_1 + iA_y\psi_1 - A_z\psi_2\right) = 0, \tag{B.27}$$

where $i\kappa$ is some coefficient. Let us rearrange the terms in expression (B.26)

$$\frac{1}{c}J^k A_k = \left(iA_y\psi_4^* + A_z\psi_3^* + A_x\psi_4^*\right)\psi_1 + \left(A_x\psi_3^* - iA_y\psi_3^* - A_z\psi_4^*\right)\psi_2 + $$
$$+ \left(A_z\psi_1^* + iA_y\psi_2^* + A_x\psi_2^*\right)\psi_3 + \left(A_x\psi_1^* - iA_y\psi_1^* - A_z\psi_2^*\right)\psi_4. \tag{B.28}$$

We substitute expression (B.28) into equation (B.27)

$$\psi_1\left(\partial_0\psi_1^* + \partial_x\psi_4^* + i\partial_y\psi_4^* + \partial_z\psi_3^* + i\kappa A_x\psi_4^* - \kappa A_y\psi_4^* + i\kappa A_z\psi_3^*\right) +$$
$$+\psi_2\left(\partial_0\psi_2^* + \partial_x\psi_3^* - i\partial_y\psi_3^* - \partial_z\psi_4^* + i\kappa A_x\psi_3^* + \kappa A_y\psi_3^* - i\kappa A_z\psi_4^*\right) +$$
$$+\psi_3\left(\partial_0\psi_3^* + \partial_x\psi_2^* + i\partial_y\psi_2^* + \partial_z\psi_1^* + i\kappa A_z\psi_1^* - \kappa A_y\psi_2^* + i\kappa A_x\psi_2^*\right) +$$
$$+\psi_4\left(\partial_0\psi_4^* + \partial_x\psi_1^* - i\partial_y\psi_1^* - \partial_z\psi_2^* + i\kappa A_x\psi_1^* + \kappa A_y\psi_1^* - i\kappa A_z\psi_2^*\right) +$$
$$+\psi_1^*\left(\partial_0\psi_1 + \partial_x\psi_4 - i\partial_y\psi_4 + \partial_z\psi_3 - i\kappa A_x\psi_4 - \kappa A_y\psi_4 - i\kappa A_z\psi_3\right) +$$
$$+\psi_2^*\left(\partial_0\psi_2 + \partial_x\psi_3 + i\partial_y\psi_3 - \partial_z\psi_4 - i\kappa A_x\psi_3 + \kappa A_y\psi_3 + i\kappa A_z\psi_4\right) +$$
$$+\psi_3^*\left(\partial_0\psi_3 + \partial_x\psi_2 - i\partial_y\psi_2 + \partial_z\psi_1 - i\kappa A_x\psi_2 - \kappa A_y\psi_2 - i\kappa A_z\psi_1\right) +$$
$$+\psi_4^*\left(\partial_0\psi_4 + \partial_x\psi_1 + i\partial_y\psi_1 - \partial_z\psi_2 - i\kappa A_x\psi_1 + \kappa A_y\psi_1 + i\kappa A_z\psi_2\right) = 0, \tag{B.29}$$

Let us write equation (B.29) in a quadratic form. The first four terms of equation (B.29) have the form

$$\psi_1\left(\partial_0\psi_1^* + \partial_x\psi_4^* + i\partial_y\psi_4^* + \partial_z\psi_3^* + i\kappa A_x\psi_4^* - \kappa A_y\psi_4^* + i\kappa A_z\psi_3^*\right) +$$
$$+\psi_2\left(\partial_0\psi_2^* + \partial_x\psi_3^* - i\partial_y\psi_3^* - \partial_z\psi_4^* + i\kappa A_x\psi_3^* + \kappa A_y\psi_3^* - i\kappa A_z\psi_4^*\right) +$$
$$+\psi_3\left(\partial_0\psi_3^* + \partial_x\psi_2^* + i\partial_y\psi_2^* + \partial_z\psi_1^* + i\kappa A_z\psi_1^* - \kappa A_y\psi_2^* + i\kappa A_x\psi_2^*\right) +$$
$$+\psi_4\left(\partial_0\psi_4^* + \partial_x\psi_1^* - i\partial_y\psi_1^* - \partial_z\psi_2^* + i\kappa A_x\psi_1^* + \kappa A_y\psi_1^* - i\kappa A_z\psi_2^*\right) =$$
$$= \begin{pmatrix}\vartheta^T & \eta^T\end{pmatrix}\partial_0\begin{pmatrix}\vartheta^*\\ \eta^*\end{pmatrix} + \begin{pmatrix}\vartheta^T & \eta^T\end{pmatrix}\begin{pmatrix}\sigma^{k*}\partial_k\eta^*\\ \sigma^{k*}\partial_k\vartheta^*\end{pmatrix} + i\kappa\begin{pmatrix}\vartheta^T & \eta^T\end{pmatrix}\begin{pmatrix}\sigma^{k*}A_k\eta^*\\ \sigma^{k*}A_k\vartheta^*\end{pmatrix} =$$
$$= \begin{pmatrix}\vartheta^T & \eta^T\end{pmatrix}\begin{pmatrix}I_2\partial_0 & \sigma^{k*}\partial_k + i\kappa\sigma^{k*}A_k\\ \sigma^{k*}\partial_k + i\kappa\sigma^{k*}A_k & I_2\partial_0\end{pmatrix}\begin{pmatrix}\vartheta^*\\ \eta^*\end{pmatrix}. \tag{B.30}$$

Let us transform the second four terms of expression (B.29), we obtain

$$\psi_1^*\left(\partial_0\psi_1 + \partial_x\psi_4 - i\partial_y\psi_4 + \partial_z\psi_3 - i\kappa A_x\psi_4 - \kappa A_y\psi_4 - i\kappa A_z\psi_3\right) +$$
$$+\psi_2^*\left(\partial_0\psi_2 + \partial_x\psi_3 + i\partial_y\psi_3 - \partial_z\psi_4 - i\kappa A_x\psi_3 + \kappa A_y\psi_3 + i\kappa A_z\psi_4\right) +$$
$$+\psi_3^*\left(\partial_0\psi_3 + \partial_x\psi_2 - i\partial_y\psi_2 + \partial_z\psi_1 - i\kappa A_x\psi_2 - \kappa A_y\psi_2 - i\kappa A_z\psi_1\right) +$$
$$+\psi_4^*\left(\partial_0\psi_4 + \partial_x\psi_1 + i\partial_y\psi_1 - \partial_z\psi_2 - i\kappa A_x\psi_1 + \kappa A_y\psi_1 + i\kappa A_z\psi_2\right) =$$
$$= \begin{pmatrix}\vartheta^\dagger & \eta^\dagger\end{pmatrix}\partial_0\begin{pmatrix}\vartheta \\ \eta\end{pmatrix} + \begin{pmatrix}\vartheta^\dagger & \eta^\dagger\end{pmatrix}\begin{pmatrix}\sigma^k\partial_k\eta \\ \sigma^k\partial_k\vartheta\end{pmatrix} - i\kappa\begin{pmatrix}\vartheta^\dagger & \eta^\dagger\end{pmatrix}\begin{pmatrix}\sigma^k A_k\eta \\ \sigma^k A_k\vartheta\end{pmatrix} =$$
$$= \begin{pmatrix}\vartheta^\dagger & \eta^\dagger\end{pmatrix}\begin{pmatrix}I_2\partial_0 & \sigma^k\partial_k - i\kappa\sigma^k A_k \\ \sigma^k\partial_k - i\kappa\sigma^k A_k & I_2\partial_0\end{pmatrix}\begin{pmatrix}\vartheta \\ \eta\end{pmatrix}.$$

(B.31)

Taking into account expressions (B.30) and (B.31), equation (B.29) takes the form

$$\begin{pmatrix}\vartheta^T & \eta^T\end{pmatrix}\begin{pmatrix}I_2\partial_0 & \sigma^{k*}\partial_k + i\kappa\sigma^{k*} A_k \\ \sigma^{k*}\partial_k + i\kappa\sigma^{k*} A_k & I_2\partial_0\end{pmatrix}\begin{pmatrix}\vartheta^* \\ \eta^*\end{pmatrix} +$$
$$+ \begin{pmatrix}\vartheta^\dagger & \eta^\dagger\end{pmatrix}\begin{pmatrix}I_2\partial_0 & \sigma^k\partial_k - i\kappa\sigma^k A_k \\ \sigma^k\partial_k - i\kappa\sigma^k A_k & I_2\partial_0\end{pmatrix}\begin{pmatrix}\vartheta \\ \eta\end{pmatrix} = 0.$$

(B.32)

Let us introduce the matrix operator

$$L \overset{\text{det}}{=} \begin{pmatrix}I_2\partial_0 & \sigma^k\partial_k - i\kappa\sigma^k A_k \\ \sigma^k\partial_k - i\kappa\sigma^k A_k & I_2\partial_0\end{pmatrix} =$$
$$= \begin{pmatrix}I_2 & 0 \\ 0 & -I_2\end{pmatrix}\begin{pmatrix}I_2\partial_0 & \sigma^k\partial_k - i\kappa\sigma^k A_k \\ -\sigma^k\partial_k + i\kappa\sigma^k A_k & -I_2\partial_0\end{pmatrix} = \gamma^0 \mathcal{L}.$$

(B.33)

Let us write equation (B.32) through operator (B.33)

$$\psi^T L^*\psi^* + \psi^\dagger L\psi = 0,$$
$$\Lambda \overset{\text{det}}{=} \psi^\dagger L\psi, \quad \Lambda^* + \Lambda = 0 \Rightarrow \operatorname{Re}\Lambda = 0 \Rightarrow \Lambda = iu, u \in \mathbb{R}.$$

(B.34)

Real function $u$ can be represented as a quadratic form

$$u \overset{\text{det}}{=} \psi^\dagger V\psi = \begin{pmatrix}\vartheta^\dagger & \eta^\dagger\end{pmatrix}\begin{pmatrix}\lambda_1 I_2 & 0 \\ 0 & \lambda_2 I_2\end{pmatrix}\begin{pmatrix}\vartheta \\ \eta\end{pmatrix} = \begin{pmatrix}\vartheta^\dagger & \eta^\dagger\end{pmatrix}\begin{pmatrix}\lambda_1\vartheta \\ \lambda_2\eta\end{pmatrix} = \lambda_1|\vartheta|^2 + \lambda_2|\eta|^2 \in \mathbb{R},$$

(B.35)

where $\lambda_{1,2} \in \mathbb{R}$ are some functions. Substituting representation (B.35) into equation (B.34), we obtain

$$\psi^\dagger L\psi = -iu = -i\psi^\dagger V\psi \Rightarrow \psi^\dagger \cdot (L\psi + iV\psi) = 0,$$
$$L\psi = -iV\psi.$$

(B.36)

Taking into account (B.35), (B.33), equation (B.36) in the matrix form takes the form:

$$\gamma^0 \begin{pmatrix} I_2 \partial_0 & \sigma^k \partial_k - i\kappa\sigma^k A_k \\ -\sigma^k \partial_k + i\kappa\sigma^k A_k & -I_2 \partial_0 \end{pmatrix} \begin{pmatrix} \vartheta \\ \eta \end{pmatrix} = -i \begin{pmatrix} \lambda_1 I_2 & 0 \\ 0 & \lambda_2 I_2 \end{pmatrix} \begin{pmatrix} \vartheta \\ \eta \end{pmatrix},$$

$$\begin{pmatrix} I_2 \partial_0 & \sigma^k \partial_k - i\kappa\sigma^k A_k \\ -\sigma^k \partial_k + i\kappa\sigma^k A_k & -I_2 \partial_0 \end{pmatrix} \begin{pmatrix} \vartheta \\ \eta \end{pmatrix} = -i \begin{pmatrix} \lambda_1 I_2 & 0 \\ 0 & -\lambda_2 I_2 \end{pmatrix} \begin{pmatrix} \vartheta \\ \eta \end{pmatrix},$$

or

$$\begin{pmatrix} \partial_0 \vartheta + \sigma^k \partial_k \eta - i\kappa\sigma^k A_k \eta + i\lambda_1 \vartheta \\ -\partial_0 \eta - \sigma^k \partial_k \vartheta + i\kappa\sigma^k A_k \vartheta - i\lambda_2 \eta \end{pmatrix} = \begin{pmatrix} 0 \\ 0 \end{pmatrix},$$

$$\partial_0 \begin{pmatrix} I_2 & 0 \\ 0 & -I_2 \end{pmatrix} \begin{pmatrix} \vartheta \\ \eta \end{pmatrix} + \begin{pmatrix} 0 & \sigma^k \partial_k \\ -\sigma^k \partial_k & 0 \end{pmatrix} \begin{pmatrix} \vartheta \\ \eta \end{pmatrix} - i\kappa \begin{pmatrix} 0 & \sigma^k A_k \\ -\sigma^k A_k & 0 \end{pmatrix} \begin{pmatrix} \vartheta \\ \eta \end{pmatrix} + i \begin{pmatrix} \lambda_1 I_2 & 0 \\ 0 & -\lambda_2 I_2 \end{pmatrix} \begin{pmatrix} \vartheta \\ \eta \end{pmatrix} = \begin{pmatrix} 0 \\ 0 \end{pmatrix}$$

(B.37)

Let us make a change of variables

$$\lambda_{1,2} \stackrel{\text{det}}{=} \frac{1}{\hbar c}(U \pm mc^2),$$

(B.38)

where $\hbar, c, m$ are some constant values, and $U \in \mathbb{R}$ is a function. Substituting (B.38) into equation (B.37), we

$$\begin{pmatrix} \lambda_1 I_2 & 0 \\ 0 & -\lambda_1 I_2 \end{pmatrix} \begin{pmatrix} \vartheta \\ \eta \end{pmatrix} = \frac{U}{\hbar c} \begin{pmatrix} I_2 & 0 \\ 0 & -I_2 \end{pmatrix} \begin{pmatrix} \vartheta \\ \eta \end{pmatrix} + \frac{mc^2}{\hbar c} \begin{pmatrix} I_2 & 0 \\ 0 & I_2 \end{pmatrix} \begin{pmatrix} \vartheta \\ \eta \end{pmatrix} = \frac{U}{\hbar c} \gamma^0 \psi + \frac{mc^2}{\hbar c} I \psi$$

or

$$i\hbar c \partial_0 \gamma^0 \psi + i\hbar c \gamma^k \partial_k \psi + \hbar c \kappa \gamma^k A_k \psi - U\gamma^0 \psi - mc^2 I \psi = 0,$$
$$i\hbar c \gamma^0 \partial_0 \psi = \left[ c\gamma^k (\hat{p}_k - \hbar\kappa A_k) + U\gamma^0 + mc^2 I \right] \psi.$$

(B.39)

If $U = q\varphi$ and $\hbar\kappa = q$, then the equation coincides with (B.39) and will take the form

$$i\hbar c \partial_0 \psi = \gamma^0 \left[ c\gamma^k (\hat{p}_k - q A_k) + mc^2 I \right] \psi + q\varphi\psi.$$

(B.40)

Let us write equation (B.40) in another form

$$i\hbar c (\gamma^0 \partial_0 + \gamma^k \partial_k) \psi - qc \left( \gamma^0 \frac{\varphi}{c} - \gamma^k A_k \right) \psi - mc^2 I \psi = 0,$$
$$i\hbar c \gamma^\mu \partial_\mu \psi - qc (\gamma^0 A_0 + \gamma^k A_k) \psi - mc^2 I \psi = 0,$$
$$\left[ \gamma^\mu (i\hbar \partial_\mu - qA_\mu) - mc I \right] \psi = 0,$$

(B.41)

where the relation $A_k = -A_k$ between the covariant components of four-vector $A_k$ and the components of 3D vector $A_k$ is taken into account. Theorem 4 is proved.